\documentclass[aps,prl,floatfix,epsfig,twocolumn,showpacs,preprintnumbers]{revtex4}
\usepackage{amsmath}
\usepackage{graphicx}
\usepackage{epstopdf}
\usepackage{dcolumn}
\usepackage{bm}
\usepackage{mathrsfs}
\usepackage[pdfstartview=FitH, CJKbookmarks=true, bookmarksnumbered=true, bookmarksopen=true, pdfborderstyle={/S/U}, pdfborder={0 0 1}, colorlinks=true, linkcolor=blue, anchorcolor=green, citecolor=blue]{hyperref}

\input{tcilatex}
\begin{document}

\title{Turning Copper Metal into Weyl Semimetal}
\author{Yongping Du}
\affiliation{Department of Applied Physics and Institution of Energy and Microstructure
Nanjing University of Science and Technology, Nanjing, Jiangsu 210094, P. R.
China.}
\author{Er-jun Kan}
\affiliation{Department of Applied Physics and Institution of Energy and Microstructure
Nanjing University of Science and Technology, Nanjing, Jiangsu 210094, P. R.
China.}
\author{Hu Xu}
\affiliation{Department of Physics, South University of Science and Technology of China,
518055 Shenzhen, China}
\author{Sergey Y. Savrasov}
\email{savrasov@physics.ucdavis.edu}
\affiliation{Department of Physics, University of California, Davis, CA 95616, USA}
\author{Xiangang Wan}
\email{xgwan@nju.edu.cn}
\affiliation{National Laboratory of Solid State Microstructures and Department of
Physics, Nanjing University, Nanjing 210093, China}
\affiliation{Collaborative Innovation Center of Advanced Microstructures, Nanjing
University, Nanjing 210093, China}

\begin{abstract}
A search for new topological quantum systems is challenging due to the
requirement of non--trivial band connectivity that leads to protected
surface states of electrons. A progress in this field was primarily due to a
realization of band inversion mechanism between even and odd parity states
that was proven to be very useful in both predicting many of such systems
and our understanding their topological properties. Despite many proposed
materials assume the band inversion between $s$ and $p$ (or $p/d$, $d$/$f$)
electrons, here, we explore a different mechanism where the occupied $d$
states subjected to a tetrahedral crystal field produce an active $t_{2g}$
manifold behaving as a state with an effective orbital momentum equal to
negative one, and pushing $j_{eff}=1/2$ doublet at a higher energy. Via
hybridization with nearest neighbor orbitals realizable, \textit{e.g.,} in a
zincblende structural environment, this allows a formation of odd parity
state whose subsequent band inversion with an unoccupied s band becomes
possible, prompting us to look for the compounds\ with Cu$^{+1}$ ionic
state. Chemical valence arguments coupled to a search in materials database
lead us to systematically investigate electronic structures and topological
properties of CuY (Y=F, Cl, Br, I) and CuXO (X=Li, Na, K, Rb) families of
compounds.\ Our theoretical results show that CuF displays a behavior
characteristic of an ideal Weyl semimetal with 24 Weyl nodes at the bulk
Brillouin Zone. We also find that another compounds CuNaO and CuLiO are the $%
s$--$d$ inversion type topological insulators. Results for their electronic
structures and corresponding surfaces states are presented and discussed in
the context of their topological properties.
\end{abstract}

\date{\today }
\maketitle

\section{Introduction}

Topological quantum solids are a new class of systems that behave as an
insulator or semimetal in the bulk but whose surface contains conducting
states meaning that the electrons can primarily move along the surface of
the material. Starting from the original proposal on the principal existence
of such state of matter in the case of two dimensions (2D) called a quantum
spin Hall insulator \cite{QSHI,QSHI-BHZ} and its subsequent extension to its
three--dimensional (3D) analog called a topological insulator (TI) \cite%
{3DTI}, the field has been recently enriched by the discoveries of new
topological phases of matter such as topological crystalline insulators \cite%
{TCI}, Weyl semimetals (WSM) \cite{WSM}, Dirac semimetals (DSM) \cite%
{DSM,DSM2,DSM3}, and nodal line semimetals (NLS) \cite{NLS}. Their unusual
properties such as robust surface currents insensitive to the disorder or
various forms of quantum Hall effects have led to a plethora of proposals
for the use of topological materials in fundamental research spanning from
magnetic monopoles to Majorana fermions, and in applications such as
fault--tolerant quantum computations \cite{RMPTI,RMPWSM}.

How do we identify new topological materials in the infinite world of
chemically allowed compounds? Although, group theoretical arguments have
been recently shown to be very helpful in our understanding of band
connectivity and their resulting topological behavior \cite{Vish1,2D-case},
the remarkable progress in this field has been primarily driven by a
realization of the so called band inversion mechanism \cite{QSHI-BHZ} where
the $s$ level at the conduction--band edge and the $p$ level at the
valence--band edge interchange at a certain portion of the Brillouin Zone
(BZ) and open a hybridization gap. Since $s$ and $p$ are even or odd parity
states, this may result in a non--zero $Z_{2}$ topological invariant
according to the Fu--Kane criterion of the 3D TIs \cite{Z2}. Much celebrated
discoveries of such topological systems as HgTe quantum wells \cite{QSHI-BHZ}%
, Bi$_{2}$Se$_{3}$ and related compounds\cite{Bi2Se3Zhang,Bi2Se3Hasan},
proposals for half--Heusler ternary compounds \cite{HalfH, HalfZhang,HalfYao}%
, BaBiO$_{3}$ \cite{BaBiO3}, \textit{etc.}, have been all based on this band
inversion mechanism. Very interesting extensions of these ideas for the
inversion between $d$ and $f$ levels have led to the proposals of some
Americium and Plutonium compounds \cite{AmC} showing non--trivial
topological properties as well as to the discovery of a topological Kondo
insulator SmB$_{6}$ \cite{SmB6}. Another example is the most recent proposal
of nodal line materials that have the $p-d$ inversion of levels \cite%
{pdinv1,pdinv2,YP1,YP2}.

\begin{figure}[tbp]
\includegraphics[scale=0.5]{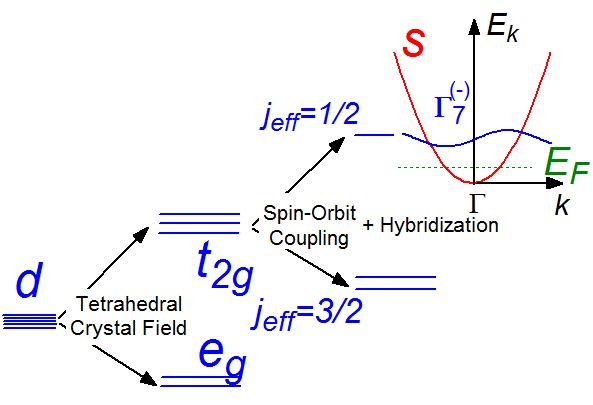}
\caption{Illustration how the atomic d--level subjected to a tetrahedral
crystal field spin--orbit coupling and hybridization with nearest neighbors
produces odd parity $\Gamma _{7}^{(-)}$ doublet that can hybridize with a
predominantly empty s--band of a material. }
\label{FigLevel}
\end{figure}

Will systems with hybridized $s$ and $d$ electrons qualify to show
topologically protected states? Despite the corresponding spherical
harmonics are both of even parity, it is well known that the $d$ states
subjected to the cubic crystal field produce a $t_{2g}$ triplet that behaves
in many cases as a state with effective orbital momentum equal to negative
one \cite{Kanamori}. Most notable examples here are the Mott insulating
behavior in Sr$_{2}$IrO$_{4}$ \cite{Sr2IrO4} and predicted non--collinear
magnetism in pyrochlore iridates \cite{WSM} where the spin--orbit splitting
of the $t_{2g}$ manifold leads to an active $j_{eff}=1/2$ doublet state. One
can subsequently explore a new class of materials where active $s$ and $d$
electrons realize new topologically non--trivial systems. This is
schematically illustrated in Fig. \ref{FigLevel} where the atomic $d$--level
subjected to a tetrahedral crystal field realizable, \textit{e.g.,} in a
zincblende structural environment, and spin--orbit coupling can form an odd
parity $\Gamma _{7}^{(-)}$ state via hybridization with nearest--neighbor
orbitals. A subsequent band inversion with an unoccupied s band becomes
possible in compounds\ containing ionic configurations d$^{10}$s$^{0},$ such
as Cu$^{+1}$ .

Here we report on a theoretical study how such $s$--$d$ inversion mechanism
is capable to produce topologically non--trivial electronic phases and lead
to new realizations of Weyl semimetals and topological insulators. First, we
use a low--energy $k\cdot p$ model to illustrate the parameter space
responsible for the appearance of various topological phases. Second, using
a valence argument, we establish a chemical space of relevant compounds, and
by coupling first--principles band structure calculations with a search in
materials database, we successfully predict that CuF is a Weyl semimetal
with 24 Weyl points in the bulk Brillouin Zone. Remarkably, this material
does not show any other trivial states crossing the Fermi level contrary to
recently discovered TaAs\cite{XiDaiTaAs,HasanTaAs} and other WSMs\cite%
{RMPWSM}. Furthermore, we also predict that CuNaO and CuLiO are the $s-d$
inversion type topological insulators.

\begin{figure}[tbp]
\includegraphics[scale=0.3]{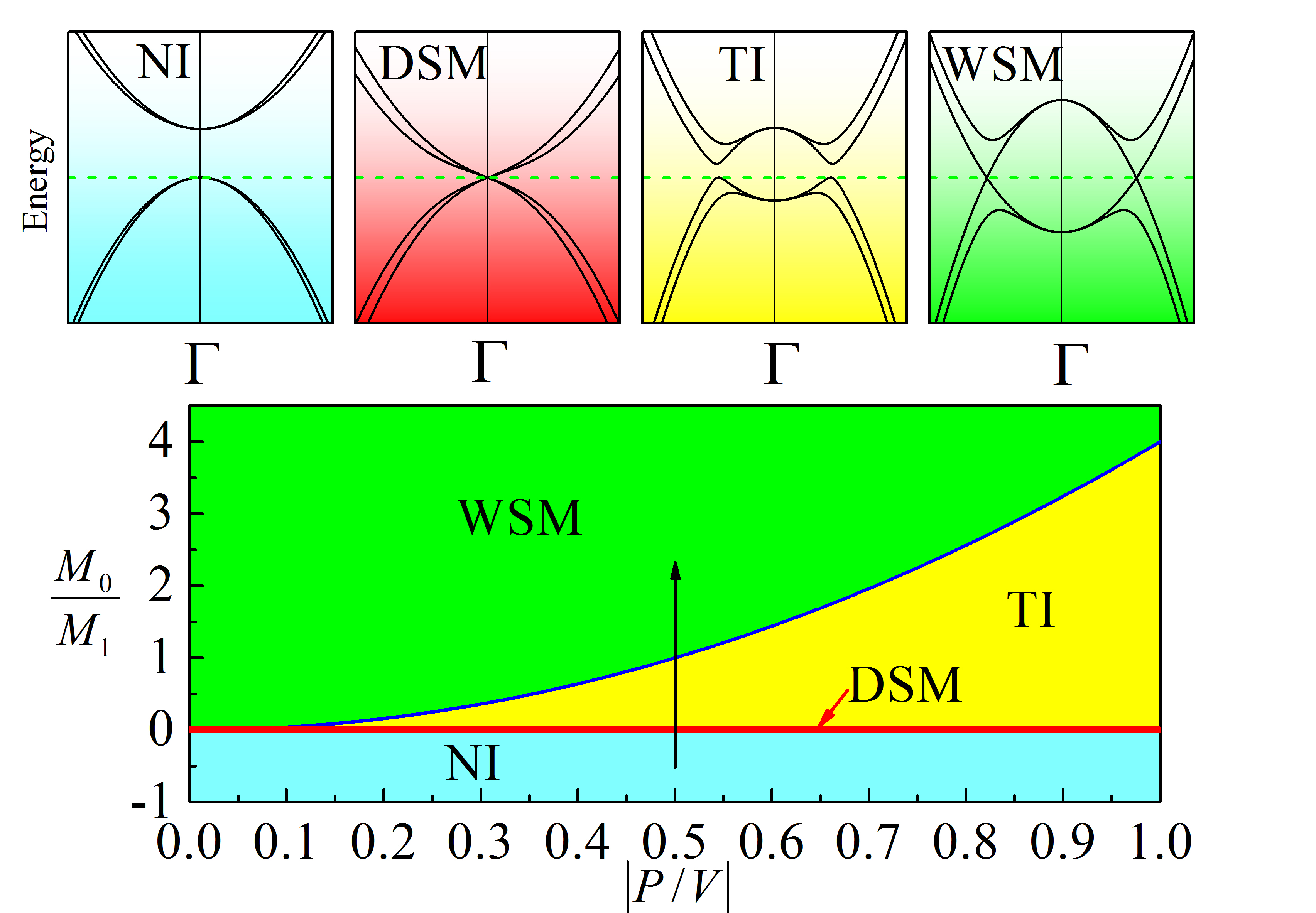}
\caption{Phase diagram of the minimal (4x4) $k\cdot p$ model involving two $%
s $--orbitals and two $\Gamma _{7}^{(-)}$ orbitals in the symmetry $%
T_{d}^{2} $ ($F\overline{4}3m$) of the zincblende structure which
illustrates realization of various topological phases in the parameter space
of the model: normal insulator (NI) (cyan), Dirac semimetal (DSM) (red),
topological insulator (TI) (yellow),Weyl semimetal (WSM) (green).}
\label{FigKP}
\end{figure}

\section{Results and Discussion}

\textbf{Effective Model.} The $d$--orbital subjected to a crystal field of
cubic lattices is split onto $t_{2g}$ and $e_{g}$ orbital states, where, if
the local symmetry is octahedral, the energy of $e_{g}$ state comes higher
than the one of $t_{2g}$ state, while if the local symmetry is tetrahedral,
the energy of the $t_{2g}$ is higher than the one of the $e_{g}$. Within the 
$t_{2g}$ manifold, the effective orbital angular momentum is one with an
additional minus sign\cite{Kanamori}. That means that when the spin--orbit
coupling (SOC)\ of the strength $\lambda $ is taken into account, the $%
t_{2g} $ state is split into effective $j_{eff}=1/2$ doublet state with and
energy $\lambda $ and $j_{eff}=3/2$ quadruplet state with an energy $%
-\lambda /2$. Note that the $j_{eff}=1/2$ state is energetically higher than
the $j_{eff}=3/2$.\emph{\ }Allowing hybridization with nearest neighbor
orbitals the $j_{eff}=1/2$ doublet can form an antibonding state of odd
parity, $|\Gamma _{7}^{(-)}\rangle $. Thus, as we illustrate on Fig. \ref%
{FigLevel}, if a predominantly empty band of the even parity ($s$) character
resides above the occupied $d$--level, its bottom around the $\Gamma $ point
and odd parity $\Gamma _{7}^{(-)}$ state can interchange and open a
hybridization gap. This scenario may produce a non--zero $Z_{2}$ topological
invariant because according to Fu and Kane criterion \cite{Z2}, the product
of occupied states parities at all time reversal invariant momenta becomes
equal to $-1$ due to the occurrence of the band inversion around the $\Gamma 
$ point.

Although many lattice space groups would qualify for exploring this idea,
the ones without inversion symmetry are particularly interesting since they
are capable of realizing Weyl semimetals. These materials were recently
proposed to have topological surface states in a form of Fermi arcs\cite{WSM}
as well as a great number of other exotic phenomena such as a highly
anisotropic negative magnetoresistance related to chiral anomaly effect\cite%
{Chiral-1,Chiral-2}, a topological response\cite{Weyl-response}, unusual
non--local transport properties\cite{non-local}, novel quantum oscillations
from the Fermi arcs\cite{Quantum-oscillations}, \textit{etc} \cite{RMPWSM}.

One of the simplest structures that would fit the requirement of the absent
inversion center is the zincblende structure with the space group $T_{d}^{2}$
($F\overline{4}3m$). Let us use a $k\cdot p$ method and construct a low
energy effective model around the $\Gamma $ point in the symmetry of the
zincblende structure by considering only a minimal basis set of 4 orbitals: $%
\left\vert s,j_{z}=1/2\right\rangle $, $\left\vert s,j_{z}=-1/2\right\rangle 
$, $\left\vert \Gamma _{7}^{(-)},j_{z}=1/2\right\rangle $, $\left\vert
\Gamma _{7}^{(-)},j_{z}=-1/2\right\rangle $.

\begin{widetext}
The Hamiltonian reads
\begin{equation*}
H_{eff}=\epsilon _{0}(\mathbf{k})+\left(
\begin{matrix}
M(\mathbf{k}) & 0 & Pk_{z}+iVk_{x}k_{y} & Pk_{-}+Vk_{z}k_{+} \\
0 & M(\mathbf{k}) & Pk_{+}-Vk_{z}k_{-} & -Pk_{z}-iVk_{x}k_{y} \\
Pk_{z}-iVk_{x}k_{y} & Pk_{-}-Vk_{z}k_{+} & -M(\mathbf{k}) & 0 \\
Pk_{+}+Vk_{z}k_{-} & -Pk_{z}+iVk_{x}k_{y} & 0 & -M(\mathbf{k})%
\end{matrix}%
\right)
\end{equation*}
\end{widetext}

where $k_{\pm }=k_{x}\pm ik_{y}$ and $M(\mathbf{k})=M_{0}-M_{1}\mathbf{k}%
^{2} $ with parameters $M_{0}$, $M_{1}$ chosen to be less than zero in order
to reproduce the band inversion. In such a case, the energy dispersion is $E(%
\mathbf{k})=\epsilon _{0}(\mathbf{k})\pm \sqrt{A(\mathbf{k})^{2}+B_{\pm }(%
\mathbf{k})^{2}}$, where $A(\mathbf{k})^{2}=M(\mathbf{k})^{2}$, $B_{\pm }(%
\mathbf{k})^{2}=P^{2}\mathbf{k}%
^{2}+V^{2}(k_{x}^{2}k_{y}^{2}+k_{x}^{2}k_{z}^{2}+k_{y}^{2}k_{z}^{2})\pm 2%
\sqrt{%
P^{2}V^{2}(k_{y}^{2}(k_{x}^{2}-k_{z}^{2})^{2}+k_{x}^{2}(k_{y}^{2}-k_{z}^{2})^{2}+k_{z}^{2}(k_{x}^{2}-k_{y}^{2})^{2})%
}$. The band crossing points are given by the following conditions: $A(%
\mathbf{k})^{2}+B_{-}(\mathbf{k})^{2}=0$.

According to the conditions of the band inversion and the band crossings, we
can obtain the phase diagram with a variety of topological phases such as
DSM, WSM, and TI, depending on the parameters of the model. This is
illustrated in Fig.\ref{FigKP}. For example, let us fix the parameter $%
\left\vert \frac{P}{V}\right\vert =0.5$ and alter $\frac{M_{0}}{M_{1}},$ as
shown by a vertical arrow in Fig.\ref{FigKP}$.$ For any $\frac{M_{0}}{M_{1}}%
<0$, there is no band inversion, and the system is a normal insulator (NI),
because the $s$ band is higher than the $\Gamma _{7}^{(-)}$ band everywhere
in k space. When $\frac{M_{0}}{M_{1}}$ approaches zero, the $s$ band lowers
down, and when $\frac{M_{0}}{M_{1}}=0$, the $s$ band exactly touches $\Gamma
_{7}^{(-)}$ band at the $\Gamma $ point. The system becomes a Dirac
semimetal with four--fold degenerate Dirac point exactly at $\mathbf{k}=0$.
The DSM plays the role of a phase boundary between topologically trivial and
non--trivial phases. It is indeed a line in the phase diagram (red) for any
value of $\left\vert \frac{P}{V}\right\vert $. If we further lower the $s$
band by increasing $\frac{M_{0}}{M_{1}}$ (both $M_{0}$, $M_{1}<0)$, the band
inversion occurs. First, we obtain the TI phase, and, second, the ideal WSM
emerges. The Weyl points are confined within $k_{x}=0$, $k_{y}=0$ or $%
k_{z}=0 $ planes by a special symmetry $C_{2T}=C_{2}\cdot T$ where $C_{2}$
denotes two--fold rotation $C_{2x}$, $C_{2y}$ and $C_{2z}$ and $T$ denotes
the time--reversal symmetry\cite{C2T}.

To illustrate the transition between the TI and the WSM phases, let us set $%
k_{z}=0$. Then the condition for bands crossing, $A(\mathbf{k})^{2}+B_{-}(%
\mathbf{k})^{2}=0$, requires both $A(\mathbf{k})=0$ and $B_{-}(\mathbf{k})=0$%
, where $A(\mathbf{k})=M_{0}-M_{1}(k_{x}^{2}+k_{y}^{2})$ and $B_{-}(\mathbf{k%
})=\left\vert P\right\vert \left\vert \sqrt{k_{x}^{2}+k_{y}^{2}}\right\vert
-\left\vert V\right\vert \left\vert k_{x}k_{y}\right\vert $. $A(\mathbf{k}%
)=0 $ is the equation for a circle and $B_{-}(\mathbf{k})=0$ is a hyperbolic
equation. When $M_{0}=0,$ the circle becomes a point and the band crossing
occurs at $\mathbf{k}=0$ giving rise to the DSM phase. When $M_{0}<0$, the
band crossing is determined by the crossing between the circle and the
hyperbola. Touching between the two occurs at $\frac{M_{0}}{M_{1}}%
=4\left\vert \frac{P}{V}\right\vert ^{2}$, which is the boundary between the
TI and the WSM phases.

\textbf{Material Realization. }A realization of such model dictates the
search of a material with a fully occupied narrow $d$ band and an empty wide 
$s$ band, so that, in principle, any ionic compound with Cu$^{+1}$ valence
state subjected to the discussed symmetry constrains qualifies. We therefore
utilize a first--principles electronic structure method (for details, see
Appendix) and systematically investigate energy bands and topological
properties of the following two families of compounds: CuX (X=F, Cl, Br, I)
and CuYO (Y=Li, Na, K, Rb). Our results indicate that the compound CuF is an
ideal Weyl semimetal while CuNaO and CuLiO are both topological insulators
of the discussed $s$--$d$ type of the band inversion. 
\begin{figure*}[tbh]
\center\includegraphics[height=0.8\textwidth,width=0.8%
\textwidth]{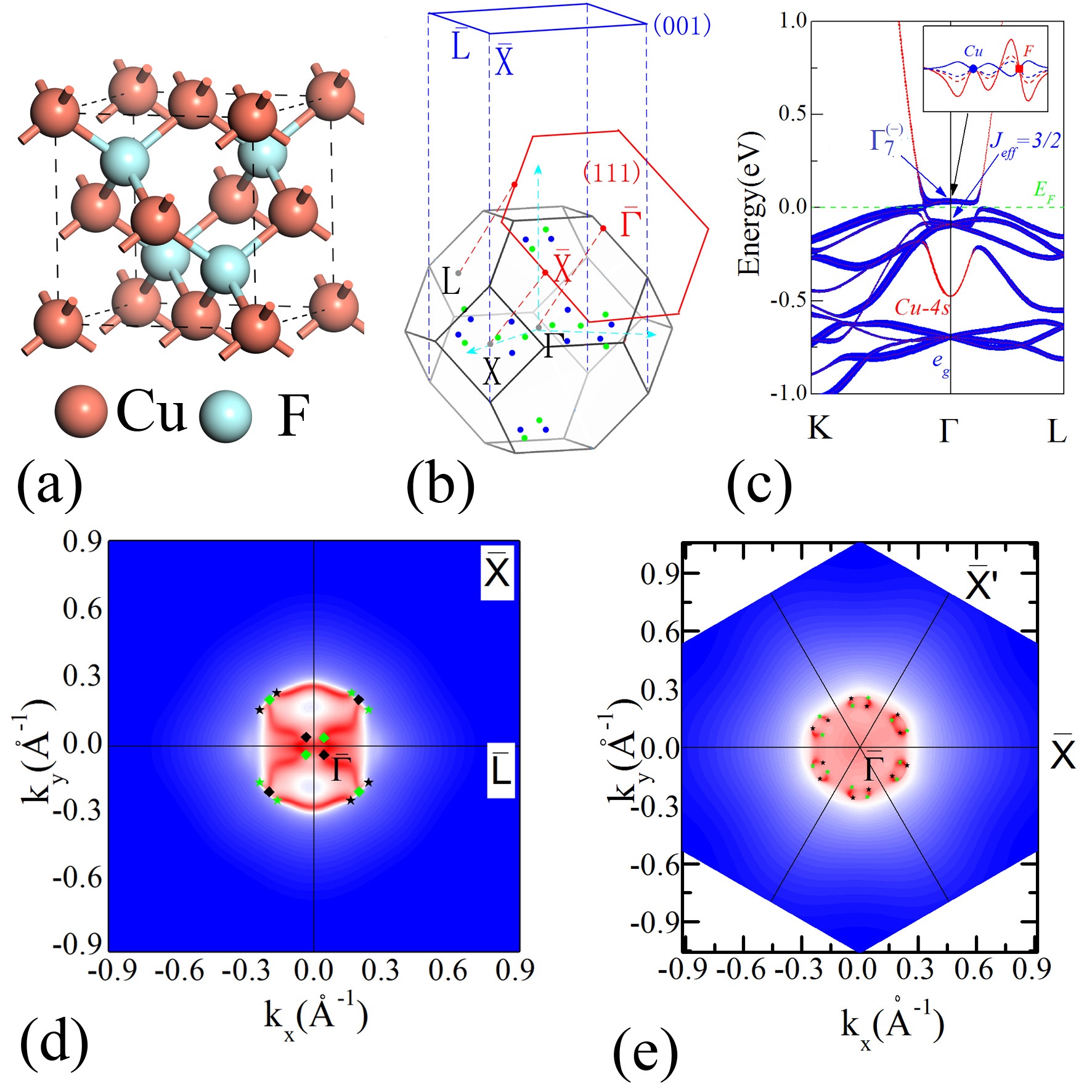}
\caption{Results for CuF: (a) Crystal structure with Cu atoms (red) and F
atoms (cyan); (b) The Brillouin Zone and the projected (001) (blue) and
(111) (red) surfaces. Green (with Chern number +1) and blue (with Chern
number -1) dots show schematic positions of the Weyl nodes. (c) Band
structure in the vicinity of the Fermi level where $\Gamma _{7}^{(-)}$
states (blue) and $s$ states (red) form band inversion. The insert shows a
wave function $|\Gamma _{7}^{(-)}$ $\rangle $\ along the Cu-F line, behaving
as an odd parity state with respect to the point approximately midway
between Cu and F. Solid/dashed lines show real (red) and imaginary (blue)
parts of two components of $\Gamma _{7}^{(-)}$, respectively. (d) The Fermi
arc surface states for the (001) surface; (e) The Fermi arc surface states
for (111) surface. Here the green (black) diamonds are the projections of
two Weyl points with monopole charges +2(--2), while the green (black) stars
are the projections of the single Weyl point with monopole charges +1(--1). }
\label{FigBandsArcsCuF}
\end{figure*}

We first discuss electronic structures of CuX (X=F, Cl, Br, I) family \cite%
{CuF}. Experimentally, it was first reported in 1933 that CuF crystallizes
in a zincblende structure\cite{CuF}, illustrated in Fig.\ref{FigBandsArcsCuF}%
(a), although its existence under ambient conditions was questioned later on%
\cite{CuF,CuF-1,CuF-2,CuF-3,CuF-4}. As there are no free internal
coordinates in this type of structure, the only parameters to optimize are
the lattice constants. Our numerical results are in a good agreement with
the original experimental work \cite{CuF} and small discrepancies have a
negligible effect on the electronic structures.

Our theoretical analysis shows that the compounds CuCl, CuBr, CuI are all
topologically trivial band insulators. The band structure of CuF along two
high symmetry lines of the first BZ (for notations, see Fig. \ref%
{FigBandsArcsCuF}(b)) is shown in Fig. \ref{FigBandsArcsCuF}(c). The
spin--orbit coupling is included in the calculation. Our orbital--character
analysis confirms that the states near the Fermi level are mainly formed by
Cu--$3d$ states (denoted by blue color) and Cu--$4s$ states (denoted by red
color) as shown in Fig. \ref{FigBandsArcsCuF}(c). F--$2p$\ states are
located between --8.5eV and --7.5eV indicating that they are far away from
the Fermi level. However, they hybridize strongly with nearest neighbor Cu--$%
j_{\emph{eff}}=1/2$\ states and produce an antibonding odd parity doublet $%
\Gamma _{7}^{(-)}$. To illustrate this, at the insert of Fig. 3(c), we plot
the calculated wave function $|\Gamma _{7}^{(-)}\rangle $ along the line
connecting Cu and F atoms. One can clearly see that, at the Cu site, the
wave function is even but in the middle of the Cu-F bond, it becomes odd via
the formation of the antibonding combination between Cu--$j_{\emph{eff}}=1/2$%
\ and F-$2p.$\emph{\ }We find that apart from the linearly dispersing Weyl
states there are no other bands crossing the Fermi level. We also find that
the $\Gamma _{7}^{(-)}$ state located at 0.04 eV\ above the Fermi energy and
the Cu--$4s$ state located at -0.47eV form an inverted band structure around
the $\Gamma $ point.

\begin{figure}[tbh]
\includegraphics[scale=0.6]{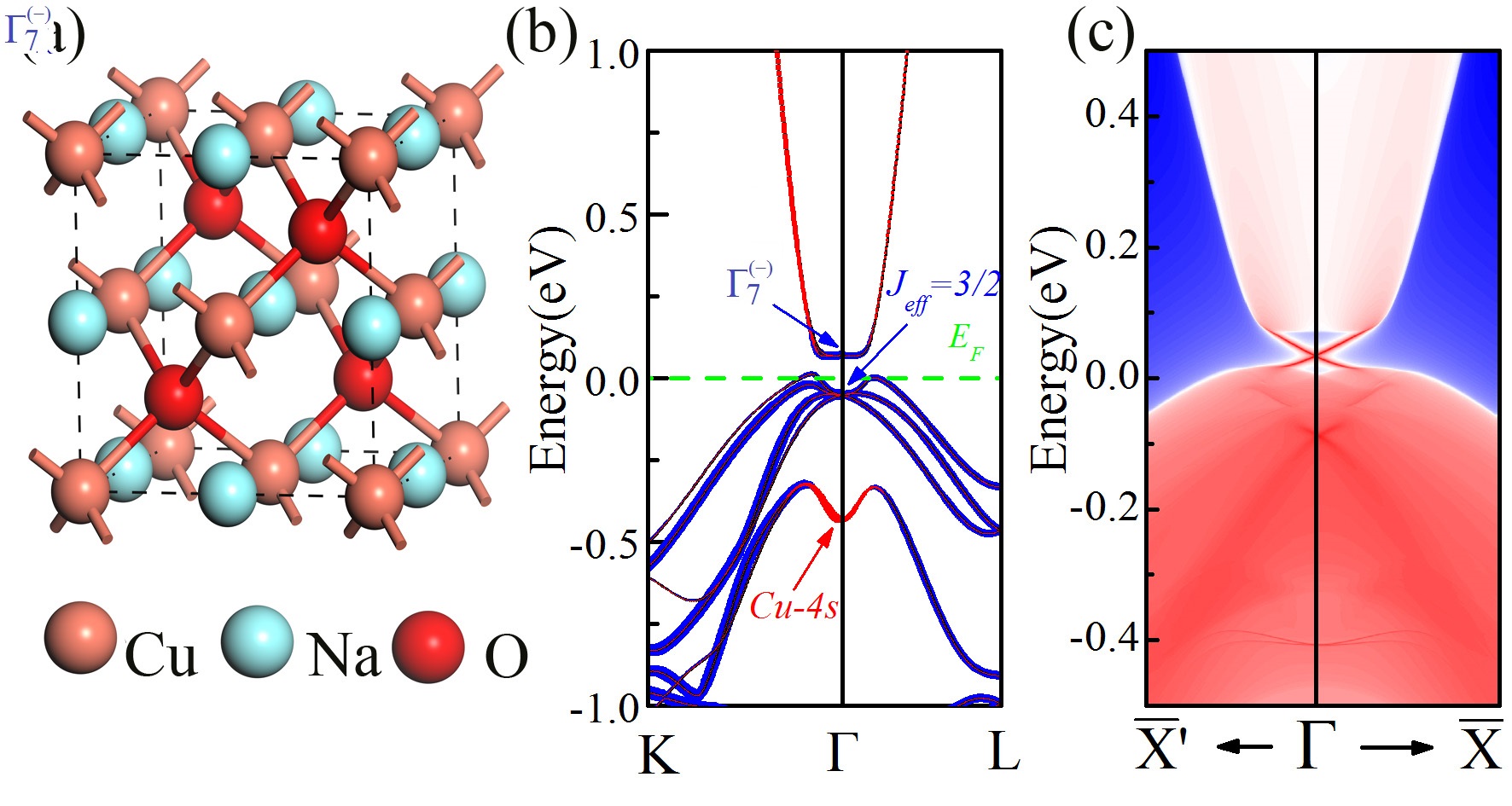}
\caption{Results for CuNaO: (a) Crystal structure where light red spheres
denote Cu atoms, dark red spheres denote O atoms while the cyan spheres
denote Na atoms; (b) The band structure in the vicinity of the Fermi level
where Cu--$j_{eff}^{ant}=1/2$ states (blue) and Cu--$4s$ states (red) form
band inversion; (c) The Dirac cone surface states for the (111) surface.}
\label{FigBandsDCsCuNaO}
\end{figure}

Our further calculation reveals that CuF is an ideal Weyl semimetal without
any additional Fermi pockets. The Weyl points are searched for by scanning
the whole BZ. To check their locations, the Chern number associated with
each Weyl point is calculated by integrating the Berry curvature based on a
computational scheme proposed by Fukui $et.al.$\cite{Chern-1}. We find that
the locations of the Weyl nodes with chirality $C=+1$ are ($\pm k_{1}$, $\pm
k_{2}$, $0$), ($0$, $\pm k_{1}$, $\pm k_{2}$), ($\pm k_{2}$, $0$, $\pm k_{1}$%
), and those with chirality $C=-1$ are ($\pm k_{1}$, $0$, $\pm k_{2}$), ( $%
\pm k_{2}$, $\pm k_{1}$, $0$), ($0$, $\pm k_{2}$, $\pm k_{1}$), where $k_{1}$%
=0.20511$\mathring{A}^{-1}$ and $k_{2}$=0.05114$\mathring{A}^{-1}$. There
are total 24 Weyl nodes related by $C_{3}$ rotational symmetry along [111]
direction and $C_{2}$ rotational symmetry in the first BZ, as shown
schematically in Fig. \ref{FigBandsArcsCuF}(b). They are confined within $%
k_{x}=0$, $k_{y}=0$ and $k_{z}=0$ planes by the co--existence of
time--reversal symmetry and the two--fold rotations, i.e. $C_{2T}$\cite{C2T}.

The existence of novel Fermi arc surface states is a remarkable feature of
Weyl semimetals. In an ideal case, such as the one we find for CuF, a well
separated set of Weyl nodes only exists at the Fermi level, and leads to
long Fermi arcs. To illustrate this, we calculate surface states of CuF by
using the Green function method based on the tight--binding Hamiltonian
obtained from maximally localized Wannier functions (MLWF) fit to the
first--principles calculation \cite{MLWF}. The Fermi arc surface states for
the (001) and (111) surfaces are shown in Fig. \ref{FigBandsArcsCuF}(d) and
Fig. \ref{FigBandsArcsCuF}(e), respectively.

In particular, for the (001) surface, the arcs are found to be long and
should be easily detectable experimentally. In Fig. \ref{FigBandsArcsCuF}%
(d), there are two touching points along $\overline{\Gamma }-\overline{X}$
which originate from the two Weyl points, with monopole charges +2 (green
diamonds) and --2 (black diamonds). Therefore, one can clearly see two
surface states coming out from these Weyl points. Along $\overline{\Gamma }-%
\overline{L}$, there are no projected Weyl points, but there exists a
surface state crossing the Fermi level which indicates a Fermi arc crossing $%
\overline{\Gamma }-\overline{L}$. The green (black) diamonds here represent
the projection of two Weyl points with monopole charges +2(--2), while the
green (black) stars represent the single Weyl point projection with monopole
charge +1(--1).

The compounds CuYO (Y=Li, Na, K, Rb) have first been reported to exist in a
tetragonal structure with space group I4/mmm\cite{CuNaO-exp1}. However, a
recent theoretical work also showed that under high hydrostatic pressure
CuYO (Y=Li, Na, K, Rb) can undergo a structural phase transition to a
zincblende structure\cite{CuNaO} which is at the center of interest in our
work. Our electronic structure calculations are performed with lattice
parameters consistent with the previous theoretical studies\cite{CuNaO,
CuNaO-2}. Our numerical results show that CuLiO and CuNaO are both
topological insulators while CuKO and CuRbO are normal insulators.

Let us take CuNaO as an example showing unique features of the $s-d$
inversion mechanism. Its crystal structure is illustrated in Fig. \ref%
{FigBandsDCsCuNaO}(a). The electronic structure with SOC included is shown
in Fig. \ref{FigBandsDCsCuNaO}(b). The active states around the Fermi level
are the Cu--$3d$ states (blue) and Cu--$4s$ states (red). From the details
at the $\Gamma $ point, the $\Gamma _{7}^{(-)}$ states are located at +0.07
eV while the $4s$ states are at -0.43eV. This illustrates the band inversion
mechanism. Our calculation of Z$_{2}$ topological invariant verifies that
this system is a TI. The corresponding band structure calculation for (111)
surface is shown on Fig.\ref{FigBandsDCsCuNaO}(c) where a linearly
dispersing Dirac cone is resolved within a small bulk energy gap.

\section{Conclusion}

In conclusion, we have explored the band inversion mechanism between
spin--orbit coupled $t_{2g}$ states and a wide $s$ band for a zincblende
type of structure. We found that a few ionic compounds with Cu$^{+1}$
valence state exhibit non--trivial topological properties: CuLiO and CuNaO
are the topological insulators while CuF is a Weyl semimetal. Remarkably,
apart from the Weyl nodes, there are no other Fermi surface states in this
compound which makes it appealing for further experimental studies of such
effects as a large negative magnetoresistance. Being a definite signature of
the chiral anomaly in the quantum limit, it has been observed in known Weyl
semimetals \cite{MagnetoR}, but it is not clear how to separate
contributions from the Weyl points and regular Fermi pockets.

Our theoretical work demonstrates how the ideas of band inversion coupled to
chemical valence considerations can guide the search of new topological
quantum systems in the infinite database of materials. Although the
tight--binding method employed in our work to detect Fermi arc surface
states does not take into account surface relaxation effects, it shows the
existence of long arcs in CuF. We have recently shown \cite{WSMDisorder},
that long and straight Fermi arcs are generally capable of supporting nearly
dissipationless surface currents, therefore it could be interesting to
explore if nanowires based on CuF are realizable in practice.

\section{Appendix: Computational Details}

We perform first--principles calculations based on the full potential
linearized augmented plane wave (FP--LAPW) method as implemented in WIEN2K
package \cite{WIEN2K}. To obtain accurate band inversion strength and band
order, the modified Becke--Johnson exchange potential together with local
density approximation for the correlation potential (MBJLDA) has been used%
\cite{mBJ}. The plane--wave cutoff parameter R$_{MT}$K$_{max}$ is set to be
7. The spin--orbit coupling is treated using the second--order variational
procedure. To check the existence of the Weyl nodes, a dense k--mesh with $%
24\times 24\times 24$ divisions \ of reciprocal lattice translations has
been employed. To study the surface states, we use the Green's function
method based on a tight--binding Hamiltonian using the maximally localized
Wannier functions (MLWFs), which are projected from the Bloch states derived
from first--principles calculations \cite{MLWF}.

\section{Acknowledgement}

The work was supported by National Key R\&D Program of China (No.
2017YFA0303203), the NSFC (No. 11525417, 51572085, 51721001 and 11790311),
National Key Project for Basic Research of China (Grant No. 2014CB921104),
the Priority Academic Program Development of Jiangsu Higher Education
Institutions. Y. P. Du was supported by the Natural Science Foundation of
Jiangsu Province (Grants No. BK20170821), the Special Foundation for
theoretical physics Research Program of China(Grants No. 11747165). S. Y. S.
was supported by NSF DMR (Grant No. 1411336).

\end{document}